\documentclass[sigconf,nonacm]{acmart}

\usepackage{hyperref}
\usepackage{subcaption}
\usepackage{graphicx}

\AtBeginDocument{%
  }

\setcopyright{acmlicensed}
\copyrightyear{2024}
\acmYear{2024}
\acmDOI{XXXXXXX.XXXXXXX}

\acmConference[HCDS '24]{3rd Workshop on Heterogeneous Composable and Disaggregated System}{April 28,
  2024}{San Diego, California, USA}
\acmISBN{978-1-4503-XXXX-X/18/06}

\begin{document}

\title{Towards Disaggregation-Native Data Streaming between Devices}

\author{Nils Asmussen}
\email{nils.asmussen@barkhauseninstitut.org}
\orcid{0000-0002-4232-4519}
\affiliation{%
  \institution{Barkhausen Institut}
  \streetaddress{Schweriner Straße 1}
  \city{Dresden}
  \country{Germany}
}

\author{Michael Roitzsch}
\email{michael.roitzsch@barkhauseninstitut.org}
\orcid{0000-0002-2416-6537}
\affiliation{%
  \institution{Barkhausen Institut}
  \streetaddress{Schweriner Straße 1}
  \city{Dresden}
  \country{Germany}
}

\renewcommand{\shortauthors}{Asmussen et al.}

\newcommand{\myos}{M\protect\raisebox{.5ex}{\protect\scalebox{.8}{3}}}

\begin{abstract}
Disaggregation is an ongoing trend to increase flexibility in datacenters.
With interconnect technologies like CXL, pools of CPUs, accelerators,
and memory can be connected via a datacenter fabric. Applications
can then pick from those pools the resources necessary for their specific
workload. However, this vision becomes less clear when we consider
data movement. Workloads often require data to be streamed through
chains of multiple devices, but typically, these data streams physically
do not directly flow device-to-device, but are staged in memory by
a CPU hosting device protocol logic. We show that augmenting devices with
a disaggregation-native and device-independent data streaming facility
can improve processing latencies by enabling data flows directly
between arbitrary devices.
\end{abstract}

\maketitle

\section{Introduction}

Many modern applications in datacenters require accelerators to run
efficiently, a major example being AI workloads requiring GPUs or
NPUs. Datacenter operators however face the challenge that not every
conceivable accelerator can be provisioned in every server as this
will lead to dramatic under-utilization of resources. Disaggregation
promises to resolve this tension by offering pools of resources
connected via a fast network fabric. Applications can select a mix
of devices fitting their workload and the underlying operating system
will configure a tailored execution environment by establishing
connections to a set of accelerators from those resource pools.

Interconnect technologies like CXL allow to disaggregate accelerators and
other devices like NICs or NVMe storage in this fashion. However,
the physical data flows pose an interesting challenge. Suppose an
application wants to pass a stream of data from the network to multiple
GPUs and NPUs for machine learning and then store the processed result
on NVMe. To reduce data movement, the shortest flow would pass data
directly from the NIC to the accelerators, where it travels between
the needed GPUs and NPUs. The last accelerator would stream its results
to the NVMe storage. In reality however, data movement will be handled
by the CPU running the application logic. Data must be staged in CPU-side
buffers to bridge between the different protocols of the involved
devices. Therefore, data will repeatedly flow between devices and the CPU
running the per-device protocol logic.

Even if CXL technically allows devices to interact directly
with each other, the different device communication protocols currently
prevent this feature from being practicable. While point solutions
like GPU access to storage~\cite{gpufs} or network~\cite{gpunet}
exist, it is unreasonable to expect every device to implement a driver
stack for every other device to enable arbitrary device-to-device
data flows. In summary, while devices are physically disaggregated
by CXL, data flows remain largely centralized due to the lack of
disaggregation-native data-movement facilities.

As a solution, we propose disaggregation-native devices, where today’s
accelerators, NICs, and storage devices are augmented with a device-independent
disaggregation-compatible data-movement facility. Towards a solution,
two questions need to be addressed: Protocol placement and inter-tenant
isolation.

\paragraph{Protocol Placement}
The system-wide data-movement facility must be device-independent,
but must interface with concrete devices. Therefore, protocol logic
that drives the system-wide device-independent facility is necessary
and must run somewhere. This logic also must interface with the
concrete devices and therefore implement device-specific logic. Depending
on the physical placement of the protocol code, the distance and therefore
the latency between application, protocol logic, and device will differ.
In this paper, we present an initial discussion of placement options for this
logic (Section~\ref{sec:data-stream}), from centralized to fully distributed.

\paragraph{Tenant Isolation}
The second challenge is the combination of direct device-to-device
data movement with a strong underlying security and isolation model.
In datacenters, tenants must only have access to specific accelerators
out of larger pools. Tenants may even space-share the same accelerator.
When such devices or device slices communicate directly with
each other, the transitive closure accessible to the tenant must still
be constrained. As data flows become less centralized and CPU-focused,
isolation enforcement becomes harder. We present requirements and
challenges for the security of device-independent data movement
(Section~\ref{sec:disaggregation-native}) and sketch a way forward.

\paragraph{Related Work}
Similar problems have been addressed in different contexts,
such as \myos{}~\cite{Asmussen:M3,Asmussen:M3x} for systems-on-chip
and FractOS~\cite{Vilanova:FractOS} for SmartNIC-based networks.
We are investigating this problem in the context of CXL. CXL is interesting,
because it combines datacenter-scale connectivity with directly attached
accelerators due to its PCIe-based nature. \myos{} also supports
accelerators directly, but only offers on-chip connectivity. FractOS
is a datacenter protocol, but relies on SmartNICs and regular servers
to attach accelerators. CXL however currently lacks our postulated device-independent
data movement facility. We quantify the potential gains of adding
disaggregation-native devices to CXL (Section~\ref{sec:evaluation}).

\section{Data Stream Protocol Placement}

\label{sec:data-stream}

\begin{figure*}[htp]
    \centering
    \begin{subfigure}{0.6\columnwidth}
        \includegraphics[width=\textwidth]{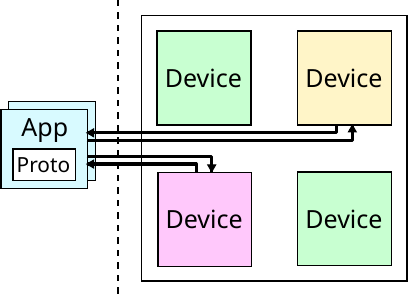}
    \end{subfigure}
    \quad
    \quad
    \quad
    \begin{subfigure}{0.6\columnwidth}
        \includegraphics[width=\textwidth]{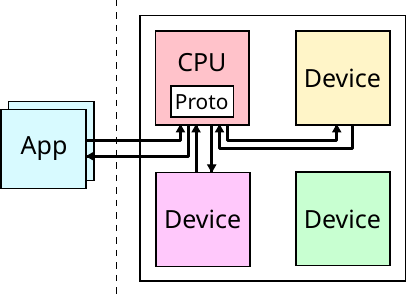}
    \end{subfigure}
    \quad
    \quad
    \quad
    \begin{subfigure}{0.6\columnwidth}
        \includegraphics[width=\textwidth]{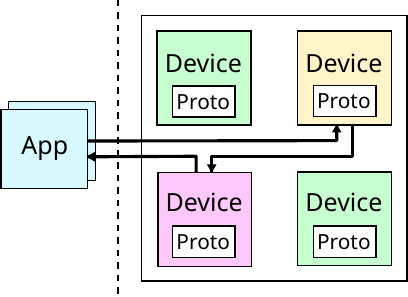}
    \end{subfigure}
    \caption{Different options for executing a protocol between application and devices:
    application-side protocol (left), central resource-side protocol (center), and distributed resource-side
    protocol (right). The dashed lines indicate machine boundaries.}
    \label{fig:protocol}
\end{figure*}

Combining multiple devices (accelerators, storage devices, network interface cards, etc.) to form
pipelines or graphs of computation requires a protocol between the application and the used devices. In
this work, we focus on the question of where such a protocol should be executed rather than how the
protocol should be designed. We consider three different options as depicted in
\autoref{fig:protocol}.

\subsection{Application-Side Protocol}

The first option constitutes the established status quo in disaggregated datacenters.
The application runs on its
machine and has access to a device pool via the CXL fabric. For simplicity, we only show
a single pool with heterogeneous devices. The devices could also be distributed among multiple pools.
With this option, the devices do not need to speak a common protocol, but can each have their own
device-specific protocol, which needs to be known by the application. In particular, the devices are not
expected to know how to talk to one another. The application therefore communicates with each device
individually using the expected protocol. Allowing applications direct access to individual devices
requires means on the device side to ensure that tenants stay within the desired boundaries and are
isolated from other tenants.

This variant therefore requires more hops during the communication in comparison to a direct
communication between devices. As depicted in \autoref{fig:protocol}, if the application wants to start
the processing on the yellow device, it needs to copy the input data to the yellow device and
trigger the start of the computation. Afterwards the application needs to first retrieve the output data
from the yellow device back to its memory and can only afterwards copy this data as input data to
the pink device. All of these steps cross the machine boundary between the application and the
resource pool, further increasing the overall latency.

\subsection{Central Resource-Side Protocol}

The second option executes the protocol within the resource pool, centrally on a CPU.
This CPU exclusively runs protocol code rather than application logic. This software
knows the individual protocols expected by the individual devices, but the
application only needs to know how to talk to the CPU. As before, the devices might be distributed among
multiple pools, each with a CPU to control them. In this case the application would need to
communicate with multiple CPUs and these in turn with their controlled devices.

Like for the application-side protocol, the central resource-side protocol does not require devices to use
a common protocol as only the CPU needs to know how to access the devices. In contrast to the
application-side protocol, applications do not access devices directly and thus the infrastructure does not
require any means to securely grant such access. Access restrictions and isolation of individual tenants
can simply be enforced by the CPU when performing device access on behalf of applications. However,
this approach still suffers in terms of latency with increasing number of devices that collaborate.
This is because the CPU still needs to access each device individually as devices cannot directly
communicate with each other. Furthermore, the CPU plays the role of an intermediary within each resource pool, also
leading to increased latency, in particular with multiple pools involved.

\subsection{Distributed Resource-Side Protocol}

The final variant executes the protocol in a distributed fashion on the devices.
Programmable accelerators such as GPUs can execute the protocol as part of the application logic.
Other devices such as SSDs already employ a processor next to the device, which can be used to
execute communication protocols. Alternatively, a processor or a fixed-function logic block can be
added next to the device to execute the protocol. If the protocol is implemented in software, it is
imaginable to allow applications to deploy the desired protocol onto the devices. Custom protocols can
provide more flexibility and/or efficiency at the cost of more complexity on the device side to
ensure isolation between tenants.

Executing the protocol on the devices themselves can lead to latency reductions as it avoids an
intermediary in the communication and can benefit from a lower latency if multiple devices are located
within the same pool. For example, as depicted in \autoref{fig:protocol}, the application directly sends
input to the yellow device, which in turn sends the output directly to the pink device, which
finally sends the result back to the application. This variant therefore leads to the least amount of
hops and keeps the communication pool-local, if possible, but requires all devices to understand
the same protocol.

\section{Disaggregation-Native Devices}

\label{sec:disaggregation-native}

Considering the different options in the previous section, we believe that the distributed
resource-side protocol is the most promising candidate as it offers the lowest possible latency. We
call devices that communicate in a peer-to-peer fashion and without CPU involvement
\emph{disaggregation-native devices}. We observe the following requirements for such devices:

\begin{enumerate}
    \item \textbf{Direct communication:} To minimize communication overhead, accelerators and
        other devices should communicate directly and avoid intermediaries such as the CPU in
        interactions. This is particularly important for longer chains of accelerators, where a
        star-shaped communication can lead to large overheads. In any case, removing intermediaries
        from interactions can lead to significant energy savings~\cite{Asmussen:M3x}.
    \item \textbf{Access restrictions:} With a CPU-centric approach, mutually distrusting tenants
        can be isolated from each other via traditional means on the CPU (e.g, different address
        spaces and memory mappings). However, direct communication between accelerators demands that
        individual accelerators can be restricted. These restrictions should be enforced externally
        instead of by the accelerator itself (e.g., via IOMMUs) to avoid that all tenants need to
        trust all accelerators, which are typically provided by third-party vendors.
    \item \textbf{Generic or custom protocols:} Combining different kinds of accelerators and
        devices requires a common protocol instead of the per-device-type protocols used
        today. This can be either achieved by designing a generic protocol for data movement that suites all desired
        use cases sufficiently well and can be implemented by all accelerators. Alternatively, it is
        imaginable that applications can bring their own protocol and load it onto the participating
        accelerators.
    \item \textbf{Protocol deployment:} The protocols should be implementable by different kinds of
        devices. For example, programmable accelerators such as GPUs might not need additional
        hardware, but can simply implement the protocol in software as part of the application
        logic. Other accelerators might already use a co-processor for internal management tasks and
        could implement the protocol on the co-processor. Fixed-function accelerators might want to
        introduce a co-processor for that purpose or implement the protocol as a fixed-function
        logic block.
    \item \textbf{Cross-machine communication:} Ideally, accelerators and devices should be able to
        interact with each other regardless of their physical location. This requires that the same
        protocol can be used even if one communication partner resides in a different machine.
        However, due to the potentially different performance characteristics, optimizing the
        placement or selection of communication partners should be feasible.
\end{enumerate}

\subsection[M3 System Architecture]{\myos{} System Architecture}

\begin{figure}
  \begin{center}
    \includegraphics[width=\columnwidth]{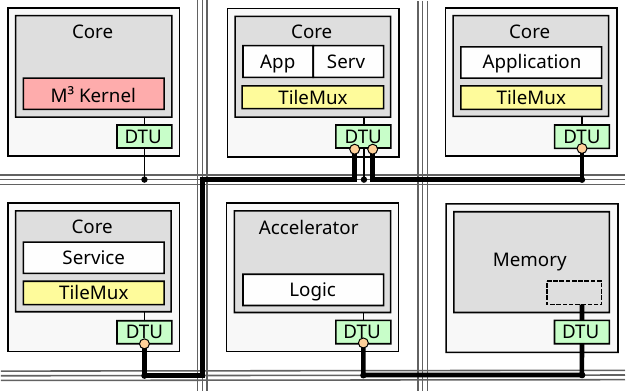}
  \end{center}
  \caption{System architecture of \myos{}: one DTU per tile isolates tiles from each other and
  selectively allows communication as configured by the \myos{} kernel. TileMux multiplexes its tile
  among the applications on this tile.}
  \label{fig:sysarch}
  \vspace{-1em}
\end{figure}

Given these requirements, we believe that the \myos{} system architecture~\cite{Asmussen:M3} is a
good starting point. Previously focused on system-on-chips~(SoCs), \myos{} builds upon a tiled
hardware architecture~\cite{wentzlaff2007chip} and runs a custom tailored operating system on top,
as shown in \autoref{fig:sysarch}. Most importantly, each tile is equipped with a new hardware
component called \emph{data transfer unit}~(DTU), which is used for cross-tile messaging and memory
accesses instead of relying on coherent shared memory.

\subsubsection{Heterogeneous Devices}

\myos{} was designed for heterogeneous systems from the beginning and thus tried to minimize the
assumptions on the individual devices. For that reason, the \myos{} kernel (red) runs on a dedicated
\emph{kernel tile} and leaves the remaining \emph{user tiles} free for applications. User tiles
therefore do not need OS support (such as virtual memory or different privilege levels), simplifying
the integration of accelerators and other devices into user tiles. Nevertheless, OS services such as
file systems and network stacks are available on all user tiles, because these are offered via
DTU-based communication protocols.

The \myos{} kernel manages the applications and OS services on user tiles as \emph{activities},
comparable to processes. An activity on a general-purpose tile executes code, whereas an activity on
an accelerator tile uses the accelerator's logic.

\subsubsection{Access Restrictions}

Supporting DTU-based communication between user tiles requires the enforcement of access
restrictions by the DTU. For that reason, message passing and memory accesses via DTU require an
established \emph{communication channel} (thick black lines in the figure). Communication channels
are represented as \emph{endpoints} in the DTU (orange dots). At runtime, each endpoint can be
configured to different endpoint types: A \emph{receive endpoint} allows to receive messages, a
\emph{send endpoint} allows to send messages to a specific receive endpoint, and a \emph{memory
endpoint} allows to issue DMA requests to tile-external memory.

Activities can use existing communication channels, but only the \myos{} kernel is allowed to
establish such channels. This is done via capabilities like in other microkernel-based
systems~\cite{fiasco,nova,sel4}. By default, no communication channels exist and thus tiles are
isolated from each other. Additionally, applications are placed on different tiles by default,
but as shown by \myos{}v~\cite{Asmussen:M3v}, tiles with general-purpose cores can also be shared
efficiently and securely among multiple applications. For that reason, every core-based user tile
runs a multiplexer called \emph{TileMux} (yellow), which is responsible for isolating and scheduling
the applications on its own tile, similar to a traditional kernel. However, in contrast to a kernel,
each TileMux instance has no permissions beyond its own tile. Instead, only the \myos{} kernel can
make system-wide decisions, hence its name.

\subsubsection{Direct Communication}

The DTU is designed to support direct communication between user tiles. For flexibility and
efficiency reasons, the DTU's interface is therefore split into a control plane and data plane. The
control plane is used during the setup phase by the \myos{} kernel to establish the required
communication channels. The data plane is used by applications to use the previously established
communication channels. During the setup phase capabilities are used to manage and distributed
permissions in the system and also to constrain the communication channels as desired. The
constraints are enforced by the DTU's data plane.

\subsubsection{Protocol Implementation}

The DTU provides message passing and DMA-like memory accesses to implement arbitrary protocols
between activities. These mechanisms have been used in the past to implement, for example, system
calls, network access, and file access. \myos{}x has also shown with the so called \emph{File
Protocol} that such protocols can be implemented both in software and as a fixed-function logic
block for simple accelerators. This design allows to combine regular applications and accelerators
in pipelines.

\subsection{Challenges}

In summary, we believe that the building blocks provided by \myos{} are well suited for
disaggregation-native devices. However, multiple challenges and open questions remain.

\begin{enumerate}
    \item \myos{} is currently designed for SoCs and therefore optimized for a low communication
        latency between tiles. Extending the system to off-chip communication with PCIe or
        cross-machine communication with CXL will increase the latency and thus require adaptions in
        hardware and software. For example, polling until a message is delivered successfully might
        no longer be desirable.
    \item The \myos{} kernel currently manages the user tiles within the same SoC from a dedicated
        kernel tile within that SoC. Moving to disaggregation-native devices that do neither need
        nor desire CPU cores raises the question whether the devices can be managed externally (e.g.,
        from a server with CPUs) by an equivalent of the \myos{} kernel.
    \item How a protocol has to be designed to be efficient and generic is an open question.
        Similarly, it is unclear whether the system should rather support the deployment of custom
        application-specified protocols instead of implementing a single generic protocol.
\end{enumerate}

\section{Evaluation}
\label{sec:evaluation}

We intend to demonstrate the benefits of a distributed resource-side protocol and
study the suitability of \myos{} as a foundation for disaggregation-native devices.
We therefore perform experiments based on the current state of \myos{}, which is available as open
source~\footnote{\url{https://github.com/Barkhausen-Institut/M3}}.

\subsection{Measurement Setup}

We use the gem5 platform~\cite{gem5} and take advantage of its customizability to configure the
system similar to future disaggregated CXL-based systems. We configure gem5 to simulate two
machines connected over an interconnect and equip each machine with multiple locally
connected devices. As the exact performance characteristics of CXL are still unknown, we
optimistically assume rather low latencies, because higher latencies further increase the latency benefit of the
distributed resource-side protocol. Concretely, we configure 1$\mu$s round-trip latency across the two
machines and 500ns round-trip latency within the machine. The latter is half of a typical
round-trip latency for PCIe~gen~3~\cite{rota2016high,erickson2018nstx}.

The application runs on a CPU in the first machine and all devices (and potential
protocol CPUs) are located in the second machine, acting as the resource pool.
Since we focus on the protocol rather than the
accelerators or devices, we do not simulate actual accelerators, but only their control cores.
We conservatively use the in-order RISC-V CPU model clocked at 1GHz for per-device control cores
and the out-of-order RISC-V CPU model with 4GHz for all other CPUs. Higher-performance control cores are of
course possible and would benefit the distributed protocol.

\subsection{Protocol Implementation}

For simplicity and comparability, we use the same protocol implementation called \emph{data channel}
for all three protocol placements. Data channels are employed between a sender and receiver
and are instantiated multiple times to form larger pipelines or graphs. The sender pushes the data into
the memory of the receiver and sends a notification message at completion. The receiver waits for this
notification and sends a response to the sender when the data has been processed. We use the data
channel to build the three placements as depicted in \autoref{fig:protocol}. For example, with
the client-side placement, the client has two data channels to the yellow device (outbound and
inbound) and two data channels to the pink device.

\subsection{Results}

\begin{figure}
    \includegraphics[width=\columnwidth]{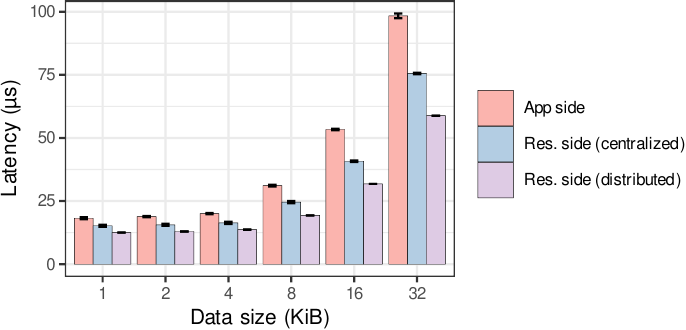}
    \caption{Performance comparison of different protocol placements using different data sizes.}
    \label{fig:results}
\end{figure}

The benchmark is run with different data sizes and uses 50 repetitions after 4 warm-up runs. As shown
in \autoref{fig:results}, the distributed version achieves the best results and is between 45\% and
67\% faster than the application-side version and 21\% to 28\% faster than the centralized version. These
results are of course preliminary due to several unknowns about future CXL-based systems. However, due
to our conservative settings we believe we can still conclude that speedups can be achieved by
executing data movement protocols in a distributed fashion on the devices themselves.

The measurements also revealed a shortcoming of the current \myos{} platform: data transfers are
performed in at most 4~KiB packets, dictated by the page size. This approach works fine on SoCs with
fast on-chip networks, but is not well suited for interconnects with higher latencies as used in
this benchmark. For this reason, the interconnect latency is paid multiple times for the data sizes
above 4~KiB leading to a faster latency increase than strictly necessary.

\section{Conclusion}

Our experiments are based on simulation, but all parameters were configured to
not unduly benefit the distributed protocol implementation. Still, we see a
significant latency benefit, leading us to conclude that disaggregation-native
devices are a logical next step in datacenter disaggregation. \myos{} has
demonstrated suitable security primitives on the system-on-chip level. It
remains open, how these primitives can be mapped to CXL fabrics and which features
of CXL must be augmented to integrate disaggregation-native devices
with strong yet decentralized inter-tenant isolation mechanisms.

\section{Acknowledgements}

We thank Mark Silberstein for the food of thought in this direction. This research is funded by the
European Union’s Horizon Europe research and innovation program under grant agreement No.~101092598
(COREnext).

\bibliography{Paper}

\end{document}